\documentclass[twocolumn,showpacs,preprintnumbers,amsmath,amssymb]{revtex4}

\usepackage{graphicx}
\usepackage{dcolumn}
\usepackage{bm}

\begin{document}


\title{
Kadowaki-Woods Ratio of
Strongly Coupled Fermi Liquids
}
\author{Takuya Okabe}

\affiliation{
Faculty of Engineering, Shizuoka University, 
3-5-1 Johoku, 
Hamamatsu 432-8561,Japan}
 \email{ttokabe@ipc.shizuoka.ac.jp}

\date{\today}

\begin{abstract}
On the basis of the Fermi liquid theory, 
the Kadowaki-Woods ratio $A/\gamma^2$ is 
evaluated by using a first principle band calculation
for typical itinerant $d$ and $f$ electron systems.
It is found as observed that the ratio for the $d$ electron systems
is significantly smaller than the normal $f$ systems,
even without considering 
their relatively weak correlation.
The difference in the ratio value 
comes from different characters of the Fermi surfaces.
By comparing Pd and USn$_3$ as typical cases,
we discuss the importance of the Fermi surface dependence
of the quasiparticle transport relaxation.
\end{abstract}

\pacs{
71.10.Ay,
71.18.+y,
71.20.Be,
71.27.+a, 
72.15.-v}
\maketitle

It is widely known as a universal feature 
of heavy fermion systems that 
there holds the Kadowaki-Woods (KW) relation
$A/\gamma^2\simeq 1\times 10^{-5} \mu\Omega$ cm(mol K/mJ)$^2$
between 
the electronic specific heat coefficient $\gamma$ 
of $C=\gamma T$
and 
the coefficient $A$ of the resistivity 
$\rho = AT^2$ in the clean and low temperature limit.\cite{KW}
According to the Fermi liquid theory,
this is interpreted as an indication of 
the fact that $A$ is squarely proportional to 
quasiparticle mass enhancement due to
strong electron correlation.
On the other hand, 
transition metal systems are 
reported since before
to obey a similar relation 
with a more than an order of magnitude smaller value of
$A/\gamma^2$.\cite{Rice,MMV}
In view of the observation that
there seems to exist several types of systems
in this regard,
the recent finding by 
Tsujii {\it et al.}\cite{TKY} is quite impressive 
that many Yb-based compounds 
show the KW ratio $A/\gamma^2$ as small as 
the transition metals.
Kontani derived the small ratio
as a result of the large orbital degeneracy of the
the $4f^{13}$ state of trivalent Yb
by applying 
the dynamical mean field approximation to
a periodic Anderson model of
an orbitally degenerate $f$ electron states
coupled with a single conduction band.\cite{HK}

To discuss 
the KW ratio $A/\gamma^2$ and 
the many-body mass enhancement effect,
a simple model is usually adopted 
at the cost of neglecting 
material specific individual factors.
%
%
In the present work, 
we are interested in such an effect 
as caused by a system-dependent factor, 
that is, 
the Fermi surface dependence 
of quasiparticle current relaxation.
The system should have a large enough Fermi surface 
relative to the Brillouin zone boundary
in order for the quasiparticle current to dissipate effectively into 
an underlying lattice
through mutual quasiparticle scatterings.
In other words, the effectiveness of the transport relaxation 
may depend on the size and shape of the Fermi surface.
To investigate this point definitely,
%
%
%
%
%
%
%
we discuss the quasiparticle transport 
by taking account of the momentum dependence of
quasiparticle scattering
on the basis of realistic band structures.
This has been hampered so far by a task required
for not so simple Fermi surfaces of many band systems 
as could be simply modelled analytically.
In terms of fairly realistic energy bands 
obtained from a first principle calculation,
we evaluate those quantities which are not affected severely by
the electron correlation effect.
The theory in use
is essentially 
within the phenomenological Fermi liquid theory
described by renormalized quantities, 
and unlike a model calculation
no bare microscopic quantities appear explicitly.
Schematic results using simple abstract models 
have been given before, in which 
a tight binding square lattice model and
a two-band model are investigated.\cite{cond1,cond2,cond3}



For the ratio $A/\gamma^2$
we make use of the expression,
\begin{align}
\frac{A}{\gamma^2} 
 = \frac{9 \alpha F}{8\pi e^2} 
 = 21.3 \alpha F a\ [\mu\Omega  \textrm{ (mol\ K/mJ)}^2],
\label{A/gamma2} 
\end{align}
which corresponds to Eq.~(4.11) in Ref.~\onlinecite{cond2}
where we set $a=4$\AA\ for the lattice constant.
In what follows we substitute a calculated value for $a$.
Below we follow how to derive $\alpha F$, 
where $\alpha$ is a coupling constant, and $F$ is 
a factor determined by the Fermi surface.


%
Following a microscopic analysis of the quasiparticle transport
with vertex corrections properly taken into account,\cite{YY}
we may derive
a phenomenological linearized Boltzmann equation.\cite{cond2}
Generalizing the theory
to take a many-band effect into account,
in the low temperature $T\rightarrow 0$
we end up with the equation
\begin{widetext}
\begin{equation}
 v^i_{p\mu} ={(\pi T)^2}
\sum_{p',k}
W^{ij}_{pp'k}
\rho^j_{p'}\rho^i_{p-k}\rho^j_{p'+k}
({l}^i_{p\mu}+{l}^j_{p'\mu}
-{l}^j_{p'+k\mu}-{l}^i_{p-k\mu}),
\label{LBeq}
\end{equation}
\end{widetext}
where
$v^i_{p\mu}$ and $\rho^i_{p} =\delta (\mu- \varepsilon^i_p)$
are the velocity component and the local density of state
of the renormalized (mass-enhanced) 
quasiparticle with the crystal momentum
$p$ in the $i$-th band.
The superscripts $i$ and $j$ are the band indices,
while the subscript $\mu={x,y,z}$ are
Cartesian coordinates.
In the right hand side of Eq.~(\ref{LBeq}),
the 2nd to 4th terms
in the parenthesis represent vertex corrections
in the microscopic formulation.
In terms of the solution ${l}^i_{p\mu}$, 
which physically represents stationary
deviation of the Fermi surface
in an applied electric field $E_\mu$,
the conductivity is given 
by
\begin{equation}
\sigma\equiv 
\sigma_{\mu}
=
2{e^2}\sum_{p, i}
\rho^i_p v^i_{p\mu} {l}^i_{p\mu},
\label{sigma0}
\end{equation}
The above equations
(\ref{LBeq}) and (\ref{sigma0}) 
correspond to Eqs.~(3.10) and  (3.15) of Ref.~\onlinecite{cond3}
respectively. 
We may suppress the index $\mu$ ($=x$) in 
Eq.~(\ref{sigma0}) as we discuss the cubic systems in what follows.

Instead of solving the simultaneous matrix equations 
(\ref{LBeq}) exactly,
we use 
trial functions for ${l}^i_{p\mu}$
as commonly applied
in a variational principle formulation of 
the transport problems.\cite{Ziman}
Assuming
\[
 l^i_{p\mu} \propto 
e^i_p \equiv \frac{v^i_{p\mu}}{|v^i_{p\mu}|},
\]
we obtain 
\begin{equation} 
\alpha F
\equiv
\frac{
\displaystyle
\sum_{i,j}
\alpha^{ij}
c_{i,j}
%
}
{\displaystyle \left(\rho^2|v_x|\right)^2
},
\label{F}
\end{equation}
where
\begin{equation}
c_{i,j}
=
\sum_{k_1,k_2,k_3,k_4 \atop k_1+k_2=k_3+k_4} 
 { \rho^i_{k_1} \rho^j_{k_2}\rho^j_{k_3} \rho^i_{k_4}
}
(e^i_{k_1}+e^j_{k_2}
-e^j_{k_3}-e^i_{k_4})^2/4 \rho_i\rho_j,
\label{cij}
\end{equation}
and 
\begin{equation}
\rho|v_x|
\equiv \sum_{i,p}
 \rho^i_p  |v^i_{px}|.
\label{rhoabsv}
\end{equation}
We define coupling constants
$\alpha^{ij} =\rho_i \rho_j 
\langle W^{ij}\rangle /\pi$,
where 
\(
\rho_i  =\sum_p \rho^i_p,
\) 
is the density of states of the $i$-th band at the Fermi level 
and
$\langle W^{ij}\rangle$ denotes
the quasiparticle scattering probability
$W^{ij}_{pp'k}$ averaged over 
the momenta $p,p'$ and $k$.
As the double sum in (\ref{LBeq}), 
dominated by Umklapp processes,
covers a complicated shaped phase space
over the Fermi surface,
it is generally a good approximation to 
take $W^{ij}_{pp'k}$
out of the momentum sum as an averaged quantity.
The total density of states 
\(
\rho =\sum_{i}\rho_i
\)
is substituted for $\gamma=2{\pi^2}\rho/{3}$.


%
In heavy fermion systems,  
the momentum dependence of $W^{ij}_{pp'k}$
could be generally neglected, for
the quasiparticle scattering $W^{ij}_{pp'k}$ is 
primarily caused by strong 
on-site Coulomb repulsion $U$.
Then we can make
an order of magnitude estimate
of $\alpha^{ii}$
in terms of 
Landau parameters $F_0^{i,s}$ and $F_0^{i,a}$.
For an anisotropic Fermi liquid, 
as in an isotropic case,
one can derive that
the charge and spin susceptibilities are given by 
$\chi_c^i =2\rho_i/(1+F^{i,s}_0)$ and
$\chi_s^i =2\rho_i/(1+F^{i,a}_0)$, respectively.
Thus, for the systems 
in which charge fluctuations are suppressed,
$\chi_c^i \rightarrow 0$,
we obtain $F^{i,s}_0\gg 1 $.
%
On the other hand, in terms of 
$A_0^{i,s}=F_0^{i,s}/(1+F_0^{i,s})$, 
one obtains a rough estimate of the coupling
\(
\alpha^{ii} =\frac{1}{4}
\left(
(A_0^{i,s}-A_0^{i,a})^2
+
\frac{1}{2}
(A_0^{i,s}+A_0^{i,a})^2
\right)
\).
Therefore, 
under the normal condition
that the spin enhancement is moderate,
$(1+F^{i,a}_0)^{-1} \sim 1$,
$\alpha^{ii}$
 should universally stay around a constant of an order of unity.\cite{cond2}
This corresponds to the condition to make
the Wilson ratio $R_W=2$ in
the impurity model.\cite{PN,YY2}
We discuss a normal state that the system is well away from 
critical instabilities, around which $A/\gamma^2$ 
will be strongly enhanced
at variance with experimental results under consideration.\cite{TM}
We evaluate $F$ numerically for $\alpha=\alpha^{ij}=1$ to obtain
$A/\gamma^2$,
and investigate the Fermi surface dependence.
It is noted that the factor $F$ is determined by the 
shape and extent of the Fermi surfaces
relative to the Brillouin zone boundary.
%
Microscopically, the mass enhancement due to 
the many-body effect is represented 
by the $\omega$-derivative of the electron self-energy $\Sigma(q, \omega)$, 
or by the renormalization factor $z_p^i$ as
$\rho^i_p =
\rho^i_{0,p}  /z_p^i$, where 
$\rho^i_{0,p}$ is a bare density of states.
It is easily checked that
the factor $z$ cancels in $F$
when $z_p^i$ is independent of $i$.
Otherwise, in case that a dominant contribution to 
the resistivity comes from an electron-correlated main band, 
then the other bands may be neglected and 
$A/\gamma^2$ becomes independent of $z$ of the main band.
As we see below numerically, 
it is found indeed that $F$ is dominated by 
a few scattering channels within a main band or two.
%
Hence, we elaborate on a numerical estimate of 
$F$ on the basis of a realistic band calculation 
reproducing reliable Fermi surfaces of 
relevant bands, even if
it may not take account of 
local many-body correlation effects fully enough
for the renormalized quantities like $\rho_i$ and $v^i_p$
to be separately compared with experiments. 
As a matter of course,
we must exclude the extreme case in which  
strong correlation 
modifies electron states around the Fermi level
qualitatively from those of a band calculation.
We apply our theory to 
those itinerant electron systems 
in which correlation strength is not negligible but not so strong.


To calculate $F$ for
some typical cubic $d$ and $f$ itinerant electron systems
in the fcc and Cu$_3$Au structures,
we have performed {\it ab initio}
band calculations 
within density functional theory using 
the plane wave pseudopotential code VASP
with the Perdew-Wang 1991 generalized gradient approximation
to the exchange correlation functional 
$E_{\rm xc}$.\cite{vasp,vasp0,vasp2,PW91}
By minimizing the total energy we obtain
the lattice constant $a$,
which is accurate enough to be used in Eq.~(\ref{A/gamma2}).

To evaluate $F$ numerically,
we have to broaden
the delta function
$\rho^i_{p} =\delta (\mu- \varepsilon^i_p)$
by $\Delta$
to pick up electron states around the Fermi level.
The width $\Delta$
of the order of real temperature
should be decreased 
as the number of the $k$-points is increased
until we confirm to have a convergent result.
For 
the number $L$ of subdivisions along reciprocal lattice vectors,
band calculations are performed with
$L_{\rm band}\sim 50$,
from which we obtain
the band energies $\varepsilon^i_k$ on
the finer $k$-mesh of $L\sim 200$
by interpolation.
As the four-fold $k$-sum
in the numerator of Eq.~(\ref{F}),
especially for the most important terms coming from
the main $d$ or $f$ correlated bands,
 constitutes
the most time consuming part of the calculation,
we have to reduce the numerical task 
by some symmetry considerations not only on 
the cubic symmetry of the quasiparticle states,
but on the relative directions of the
four momentum vectors of the scattering 
quasiparticle states
and the $x$-direction of the current flow.
The reduction is 
particularly effective for the intra-band scatterings $i=j$.

\begin{table}[t]
\caption{Calculated results.
}
\label{table}
\begin{ruledtabular}
\begin{tabular}{lccccc}
&$a$ (\AA)
&
$\rho|v_x|$~\footnotemark[1] 
& $F$ &$N$&$A/\gamma^2$~\footnotemark[2] 
\\
\hline
 USn$_3$&  4.60 
& 
3.1& 4.0&3 &0.39 \\
 UIn$_3$&  4.61 
& 
4.9& 1.6& 3&0.16\\
 UGa$_3$& 4.24 
& 
3.9& 2.5 &3& 0.23\\
\hline
 Pd& 3.86 
& 
7.4& 0.23 & 3& 0.019\\
 Pt& 3.91 
& 
8.4&0.15 &4& 0.012\\
\end{tabular} 
\end{ruledtabular}
\footnotetext[1]{In unit of $a=1$.}
\footnotetext[2]{In unit of [10$^{-5}$ $\mu\Omega$ cm (mol K/mJ)$^2$].}
\end{table}
The calculated results are 
shown in Table~\ref{table},
where $F$ and $A/\gamma^2$ 
for $\alpha=\alpha^{ij}=1$
are 
shown along with the lattice constant $a$,
the number $N$ of 
metallic bands contributing to the resistivity,
and $\rho|v_x|$ defined in Eq.~(\ref{rhoabsv}).
We find that our results explain well 
the experimental tendency of 
an order of magnitude small values of the ratio $A/\gamma^2$ 
for the transition metal systems.
As for the absolute values of the ratio,
our results are a few times smaller than observed evenly,
but the accuracy of this order should not be taken seriously here.
Among other things, 
the results indicate that
different characters of the Fermi surfaces play an important role.

\begin{figure}[t]
\begin{center}
\includegraphics{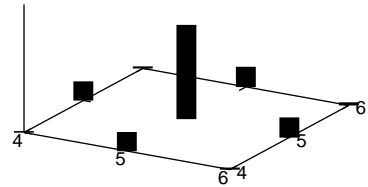}
\end{center}
\caption{
$c_{ij}$ ($i,j= 4,5,6$) for Pd.
The contribution from the 5th band is dominant for the resistivity.
}
\label{fig1}
\end{figure}

\begin{figure}[t]
\begin{center}
\includegraphics{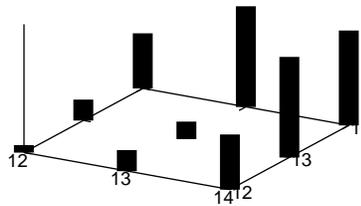}
\end{center}
\caption{
$c_{ij}$ ($i,j= 12,13,14$) for USn$_3$.
The interband contribution with 
the 14th band is important too.
}
\label{fig2}
\end{figure}

To show 
the relative contribution to the resistivity 
from relevant bands,
relative magnitudes of 
$c_{i,j}$ in the numerator of Eq.~(\ref{F})
are shown for Pd and USn$_3$
in Figs.~\ref{fig1} and \ref{fig2}, respectively.
For Pd, the contribution to $F$ 
comes from the 4th to 6th bands, among which
dominant is the 5th hole band of the $3d$ character.
Similarly, the 5th band contributes majorly not only to
$\rho$, i.e., $\rho_5\simeq 5.4 \rho_4 \simeq 12 \rho_6$,
but to $\rho|v_x|$ in Eq.~(\ref{rhoabsv}).
On the other hand, 
for USn$_3$,
while the 14th heavy electron band plays a central role, 
the 12th and 13th hole bands 
also make non-negligible contributions 
through the inter-band scatterings.
Hence, as the first point to note,
numerical importance of the inter-band contributions
makes $F$ large in the $f$ electron system.
This is partly 
because $\rho_i$ for $i=12,13,14$ are comparable
with each other, namely,
$\rho_{14}\simeq 2\rho_{13} \simeq 3 \rho_{12}$.
Moreover, it is remarked that
the large and nearly spherical shape
of the Fermi surfaces are essential too.
As the second point to note,
the importance of the Fermi surface geometry 
can be understood within a single band model
by comparing contribution from the main band.
We find that
$c_{5,5}/\rho_5^2= 0.097$ for Pd is an order of magnitude
smaller than $c_{14,14}/\rho_{14}^2= 0.93$ for USn$_3$.
The difference comes from 
the different characters of the Fermi surfaces.


According to an elementary formula
$\sigma =e^2 \rho v^2\tau
=e^2 \rho v l$,
the conductivity $\sigma$ 
depends on $\rho v$ as well as $l$.
In this context, 
the mean free path $l$ is not 
a single particle property determined by 
a lifetime of the particle state, 
but it is 
the {\it transport} property which
characterizes
how efficiently the total electric current 
decays into a lattice system, e.g., in our case,
through mutual Umklapp scattering processes between the 
current carriers.
In particular, regardless of interaction,
electrons in free space
will not have resistivity.\cite{YY}
Thus, to evaluate 
the transport property $l$ correctly, it is crucial 
to take account of the momentum dependence 
of the scattering states and their conservation 
modulo the reciprocal lattice vectors.

\begin{figure}[t]
\begin{center}
\includegraphics{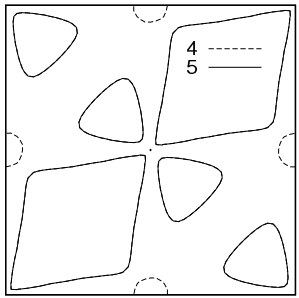}
\end{center}
\caption{
The intersection of the 
Fermi surfaces of 
Pd $d$-hole states with the (11$\bar{1}$) plane.
}
\label{fig3}
\end{figure}
\begin{figure}[t]
\begin{center}
\includegraphics{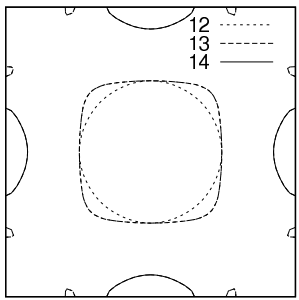}
\end{center}
\caption{
The intersection of the 
Fermi surfaces of 
USn$_3$ with the (100) plane
}
\label{fig4}
\end{figure}

Note that $\rho|v_x|$ defined in Eq.~(\ref{rhoabsv})
is related to 
the surface area $S$ of the Fermi surfaces,
as $\rho {\rm d}\varepsilon =S {\rm d}k_\perp/(2\pi)^3$.
Hence, $\rho|v_x|$ too
is independent of the mass renormalization $z$
as $F$ is, 
and for free electrons we obtain
$\rho  |v_x| \propto k_{\rm F}^2\propto n^{2/3}$.
One can see a 
correlation between $F$ and $\rho|v_x|$ 
in Table \ref{table}. 
In fact, Pd and Pt have 
twice as large $\rho|v_x|$ as the uranium compounds.
The difference cannot be simply explained 
by the difference in the Fermi surface volume $n$.
It is caused by the fact that 
the $f$-electron systems have the nearly isotropic 
Fermi surfaces while the $d$-electron systems have 
complicated ones with 
relatively large area compared to their total volume, as indicated
in Figs.~\ref{fig3} and \ref{fig4}.
The different characters of the surfaces affect
not only the single particle quantity $\rho|v_x|$ 
but also the transport property of the total current relaxation.
As 
the order of magnitude difference in $F$ is not explained 
merely by 
$\rho|v_x|$,
we have to have resort to
the other factor,
that is, 
the transport property depending on the Fermi surfaces.
It originates from 
the detailed $k$-dependence of the scattering states,
as represented in $c_{i,j}$,
or by the phase space volume available for 
all possible scattering channels
under strict restrictions of energy and momentum conservations.
%
%
%
Thus our quantitative analysis concludes the important 
effect on the quasiparticle transport due to 
the shape and complexity of the Fermi surfaces.


In summary, 
we evaluated the Kadowaki-Woods ratio $A/\gamma^2$ 
of some itinerant $d$ and $f$ electron systems
numerically on the basis of 
the Fermi liquid theory using
quasiparticle Fermi surfaces obtained by
band calculations. 
In a single framework,
we find the $d$ electron systems have smaller ratio 
than the $f$ systems, as observed,
and among others we pointed out an important effect 
to the transport coefficient $A$
originating from a commonly neglected specific feature 
depending on the characters of the Fermi surfaces.
The effect is not understood fully as a 
single-particle property of interacting systems,
but
we stress the importance of 
the phase space restriction due to
momentum conservation in two-body scattering processes 
to dissipate a total electric current.
In short, to realize effective dissipation,
the system should have a large 
and regular shaped Fermi surface. 
In future we will examine that
the Fermi-surface dependent
efficiency of mutual quasiparticle scatterings
may depend on a type of transport current to be relaxed.

\section*{Acknowledgment}

The author is grateful to N. Fujima, S. Kokado and T. Hoshino
for providing assistance in the numerical calculations.
He also acknowledges 
computational resources offered from
YITP computer system in Kyoto University.


\bibliography{okabe}

\begin{thebibliography}{17}
\expandafter\ifx\csname natexlab\endcsname\relax\def\natexlab#1{#1}\fi
\expandafter\ifx\csname bibnamefont\endcsname\relax
  \def\bibnamefont#1{#1}\fi
\expandafter\ifx\csname bibfnamefont\endcsname\relax
  \def\bibfnamefont#1{#1}\fi
\expandafter\ifx\csname citenamefont\endcsname\relax
  \def\citenamefont#1{#1}\fi
\expandafter\ifx\csname url\endcsname\relax
  \def\url#1{\texttt{#1}}\fi
\expandafter\ifx\csname urlprefix\endcsname\relax\def\urlprefix{URL }\fi
\providecommand{\bibinfo}[2]{#2}
\providecommand{\eprint}[2][]{\url{#2}}

\bibitem[{\citenamefont{Kadowaki and Woods}(1986)}]{KW}
\bibinfo{author}{\bibfnamefont{K.}~\bibnamefont{Kadowaki}} \bibnamefont{and}
  \bibinfo{author}{\bibfnamefont{S.~B.} \bibnamefont{Woods}},
  \bibinfo{journal}{Solid State Commun.} \textbf{\bibinfo{volume}{58}},
  \bibinfo{pages}{507} (\bibinfo{year}{1986}).

\bibitem[{\citenamefont{Rice}(1968)}]{Rice}
\bibinfo{author}{\bibfnamefont{M.~J.} \bibnamefont{Rice}},
  \bibinfo{journal}{Phys. Rev. Lett.} \textbf{\bibinfo{volume}{20}},
  \bibinfo{pages}{1439} (\bibinfo{year}{1968}).

\bibitem[{\citenamefont{Miyake et~al.}(1989)\citenamefont{Miyake, Matsuura, and
  Varma}}]{MMV}
\bibinfo{author}{\bibfnamefont{K.}~\bibnamefont{Miyake}},
  \bibinfo{author}{\bibfnamefont{T.}~\bibnamefont{Matsuura}}, \bibnamefont{and}
  \bibinfo{author}{\bibfnamefont{C.~M.} \bibnamefont{Varma}},
  \bibinfo{journal}{Solid State Commun.} \textbf{\bibinfo{volume}{71}},
  \bibinfo{pages}{1149} (\bibinfo{year}{1989}).

\bibitem[{\citenamefont{Tsujii et~al.}(2005)\citenamefont{Tsujii, Kontani, and
  Yoshimura}}]{TKY}
\bibinfo{author}{\bibfnamefont{N.}~\bibnamefont{Tsujii}},
  \bibinfo{author}{\bibfnamefont{H.}~\bibnamefont{Kontani}}, \bibnamefont{and}
  \bibinfo{author}{\bibfnamefont{K.}~\bibnamefont{Yoshimura}},
  \bibinfo{journal}{Phys.\ Rev.\ Lett.} \textbf{\bibinfo{volume}{94}},
  \bibinfo{pages}{057201} (\bibinfo{year}{2005}).

\bibitem[{\citenamefont{Kontani}(2004)}]{HK}
\bibinfo{author}{\bibfnamefont{H.}~\bibnamefont{Kontani}}, \bibinfo{journal}{J.
  Phys. Soc. Jpn.} \textbf{\bibinfo{volume}{73}}, \bibinfo{pages}{515}
  (\bibinfo{year}{2004}).

\bibitem[{\citenamefont{Okabe}(1998{\natexlab{a}})}]{cond1}
\bibinfo{author}{\bibfnamefont{T.}~\bibnamefont{Okabe}}, \bibinfo{journal}{J.
  Phys. Soc. Jpn.} \textbf{\bibinfo{volume}{67}}, \bibinfo{pages}{2792}
  (\bibinfo{year}{1998}{\natexlab{a}}).

\bibitem[{\citenamefont{Okabe}(1998{\natexlab{b}})}]{cond2}
\bibinfo{author}{\bibfnamefont{T.}~\bibnamefont{Okabe}}, \bibinfo{journal}{J.
  Phys. Soc. Jpn.} \textbf{\bibinfo{volume}{67}}, \bibinfo{pages}{4178}
  (\bibinfo{year}{1998}{\natexlab{b}}).

\bibitem[{\citenamefont{Okabe}(1999)}]{cond3}
\bibinfo{author}{\bibfnamefont{T.}~\bibnamefont{Okabe}}, \bibinfo{journal}{J.
  Phys. Soc. Jpn.} \textbf{\bibinfo{volume}{68}}, \bibinfo{pages}{2721}
  (\bibinfo{year}{1999}).

\bibitem[{\citenamefont{Yamada and Yosida}(1986)}]{YY}
\bibinfo{author}{\bibfnamefont{K.}~\bibnamefont{Yamada}} \bibnamefont{and}
  \bibinfo{author}{\bibfnamefont{K.}~\bibnamefont{Yosida}},
  \bibinfo{journal}{Prog. Theor. Phys.} \textbf{\bibinfo{volume}{76}},
  \bibinfo{pages}{621} (\bibinfo{year}{1986}).

\bibitem[{\citenamefont{Ziman}(1960)}]{Ziman}
\bibinfo{author}{\bibfnamefont{J.~M.} \bibnamefont{Ziman}},
  \emph{\bibinfo{title}{Electrons and Phonons}} (\bibinfo{publisher}{Clarendon
  Press}, \bibinfo{address}{Oxford}, \bibinfo{year}{1960}).

\bibitem[{\citenamefont{Nozi\`eres}(1974)}]{PN}
\bibinfo{author}{\bibfnamefont{P.}~\bibnamefont{Nozi\`eres}},
  \bibinfo{journal}{J. Low. Temp. Phys.} \textbf{\bibinfo{volume}{17}},
  \bibinfo{pages}{31} (\bibinfo{year}{1974}).

\bibitem[{\citenamefont{Yosida and Yamada}(1975)}]{YY2}
\bibinfo{author}{\bibfnamefont{K.}~\bibnamefont{Yosida}} \bibnamefont{and}
  \bibinfo{author}{\bibfnamefont{K.}~\bibnamefont{Yamada}},
  \bibinfo{journal}{Prog. Theor. Phys.} \textbf{\bibinfo{volume}{53}},
  \bibinfo{pages}{1286} (\bibinfo{year}{1975}).

\bibitem[{\citenamefont{Takimoto and Moriya}(1996)}]{TM}
\bibinfo{author}{\bibfnamefont{T.}~\bibnamefont{Takimoto}} \bibnamefont{and}
  \bibinfo{author}{\bibfnamefont{T.}~\bibnamefont{Moriya}},
  \bibinfo{journal}{Solid State Commun.} \textbf{\bibinfo{volume}{99}},
  \bibinfo{pages}{457} (\bibinfo{year}{1996}).

\bibitem[{\citenamefont{Kresse and Furthm{\"u}ller}(1996{\natexlab{a}})}]{vasp}
\bibinfo{author}{\bibfnamefont{G.}~\bibnamefont{Kresse}} \bibnamefont{and}
  \bibinfo{author}{\bibfnamefont{J.}~\bibnamefont{Furthm{\"u}ller}},
  \bibinfo{journal}{Comput. Mater. Sci.} \textbf{\bibinfo{volume}{6}},
  \bibinfo{pages}{15} (\bibinfo{year}{1996}{\natexlab{a}}).

\bibitem[{\citenamefont{Kresse and
  Furthm{\"u}ller}(1996{\natexlab{b}})}]{vasp0}
\bibinfo{author}{\bibfnamefont{G.}~\bibnamefont{Kresse}} \bibnamefont{and}
  \bibinfo{author}{\bibfnamefont{J.}~\bibnamefont{Furthm{\"u}ller}},
  \bibinfo{journal}{Phys. Rev. B} \textbf{\bibinfo{volume}{54}},
  \bibinfo{pages}{11169} (\bibinfo{year}{1996}{\natexlab{b}}).

\bibitem[{\citenamefont{Kresse and Joubert}(1999)}]{vasp2}
\bibinfo{author}{\bibfnamefont{G.}~\bibnamefont{Kresse}} \bibnamefont{and}
  \bibinfo{author}{\bibfnamefont{D.}~\bibnamefont{Joubert}},
  \bibinfo{journal}{Phys. Rev. B} \textbf{\bibinfo{volume}{59}},
  \bibinfo{pages}{1758} (\bibinfo{year}{1999}).

\bibitem[{\citenamefont{Perdew et~al.}(1992)\citenamefont{Perdew, Chevary,
  Vosko, Jackson, Pederson, Singh, and Fiolhais}}]{PW91}
\bibinfo{author}{\bibfnamefont{J.~P.} \bibnamefont{Perdew}},
  \bibinfo{author}{\bibfnamefont{J.~A.} \bibnamefont{Chevary}},
  \bibinfo{author}{\bibfnamefont{S.~H.} \bibnamefont{Vosko}},
  \bibinfo{author}{\bibfnamefont{K.~A.} \bibnamefont{Jackson}},
  \bibinfo{author}{\bibfnamefont{M.~R.} \bibnamefont{Pederson}},
  \bibinfo{author}{\bibfnamefont{D.~J.} \bibnamefont{Singh}}, \bibnamefont{and}
  \bibinfo{author}{\bibfnamefont{C.}~\bibnamefont{Fiolhais}},
  \bibinfo{journal}{Phys. Rev. B} \textbf{\bibinfo{volume}{46}},
  \bibinfo{pages}{6671} (\bibinfo{year}{1992}).

\end{thebibliography}
%
%
%
%
%
%
%
%
%
%
%
%

\end{document}